\documentclass[amsmath,amssymb,showkeys, showpacs, superscriptaddress]{revtex4}
\usepackage{graphicx}
\usepackage{amsmath,tikz}
\usetikzlibrary{matrix, positioning}
\usepackage{amsthm}%
\usepackage{amssymb}
\DeclareMathAlphabet{\mathpzc}{OT1}{pzc}{m}{it}
 \usepackage{xcolor}

\theoremstyle{remark}

\theoremstyle{definition}

\renewcommand{\a}{\alpha}

\renewcommand{\l}{\lambda}

\renewcommand{\t}{\theta}

 \newcommand{\bea}{\begin{eqnarray}} 
\newcommand{\eea}{\end{eqnarray}}  
\newcommand{\bg}{\begin{gathered}}
\newcommand{\eg}{\end{gathered}}
\newtheorem{thm}{Theorem}[section]
\newcommand{\vs}{\vskip0.1true in \noindent }

\begin{document}



\title{ The Asymptotic Iteration Method Revisited}

\author{Mourad E. H.  Ismail}
\email{mourad.eh.ismail@gmail.com. } 
\affiliation{Department of Mathematics, University of Central Florida, Orlando, Florida 32816, USA}

\author{Nasser Saad\footnote{Corresponding Author: nsaad@upei.ca }}
\affiliation{School of Mathematical and Computational Sciences,
University of Prince Edward Island, 550 University Avenue,
Charlottetown, PEI, Canada C1A 4P3.}

\begin{abstract}
\noindent The Asymptotic Iteration Method (AIM) is a technique for solving analytically and approximately the linear second-order differential equation, especially the eigenvalue problems that frequently appear in theoretical and mathematical physics. The analysis and mathematical justifications of the success and failure of the asymptotic iteration method are detailed in this work.  A theorem explaining why the asymptotic iteration method works for the eigenvalue problem is presented.  As a byproduct, a new procedure to generate unlimited
 classes of exactly solvable differential equations is also introduced.
 \end{abstract}


\maketitle

\section{Introduction}

\noindent Since the first publication \cite{Cif:Hal:Saa} of the 
Asymptotic Iteration Method (AIM), it has enjoyed great success in 
many areas of physics, among them we refer the reader to the references \cite{cho:cor:dou,cho:dou:nay,boz:bon,Ci:Ha:Sa,Dm:Mm:Jk,Gk:Ob:Ib}. 
Given the differential equation
\begin{equation}\label{eqI.1}
y''(x)=\lambda_0(x) y'(x)+s_0(x) y(x),
\end{equation}
where $\lambda_0(x)$ and $s_0(x)$ are $C^\infty$ 
functions. AIM uses the invariant
structure of the right-hand side of (\ref{eqI.1}) under further
differentiation. Indeed, the $n-1$ and $n$ differentiation of Eq.\eqref{eqI.1} gives, respectively, 
\begin{equation}\label{eqI.2}
\begin{gathered}
y^{(n+1)}(x)=\lambda_{n-1}(x)y^\prime(x)+s_{n-1}(x)y(x), 
\\
y^{(n+2)}(x)=\lambda_{n}(x)y^\prime(x)+s_{n}(x)y(x),
\end{gathered}
\end{equation}
where the AIM sequences $\lambda_n(x)$ and $s_n(x)$, $n=1,2,\dots$ are, recursively, evaluated using the relations
\begin{equation}\label{eqI.3}
\begin{gathered}
\lambda_{n}(x)=
\lambda_{n-1}^\prime(x)+s_{n-1}(x)+\lambda_0(x)\lambda_{n-1}(x)
\\
 s_{n}(x)=s_{n-1}^\prime(x)+s_0(x)\lambda_{n-1}(x).
 \end{gathered}
\end{equation}
When \eqref{eqI.1} has a polynomial solution of degree $n$, 
Ciftci, Hall, and Saad \cite{Cif:Hal:Saa} proved that the general solution  is given by 
\begin{align}\label{eqI.4}
y(x) &= C_1 \exp\left(-\int^x \a_n(t) dt\right) + C_2 \exp\left(-\int^x \a_n(t) dt\right)
\int^x \exp\left(\int^t [\l_0(u) + 2\,\a_n(u) ]du\right) dt,
\end{align}
where 
\bea
\a_n(x) =s_{n-1}(x)/\l_{n-1}(x)
\label{eqII.5}
\eea
 and $\{\lambda_k(x)\}$ and  $\{s_k(x)\}$ satisfy the termination condition
\bea
\label{eqI.6}
\delta_n(x)=\lambda_n(x)s_{n-1}(x)-\l_{n-1}(x)s_n(x)=0,\qquad n=1,2,\dots.
\eea
Using (\ref{eqI.2}), it easily follows that
\begin{equation}\label{eqI.7}
\lambda_n(x) y^{(n+1)}(x)- \lambda_{n-1}(x)y^{(n+2)}(x) = \delta_n(x)y(x) 
\end{equation}
whence if $y(x)$, the solution of
\eqref{eqI.1}, is a polynomial of degree at most $n$ then $\delta_n(x)\equiv
0$. On the other hand, if $\delta_n(x)\equiv 0$,  a particular solution of the 
differential equation \eqref{eqI.1}  is given by   $y(x) 
= \exp\left(-\int^x \a_n(t) dt\right)$. Therefore, using \eqref{eqI.2},
\bea
\label{eqI.8}
 y^{(n+1)}(x)=\left[\lambda_n(x)s_{n-1}(x) 
 -\lambda_{n-1}(x)s_n(x)\right]y(x) =0,
\eea
and consequently $y(x)$ is a polynomial of degree at most $n$. 
It is not difficult \cite{Saa:Hal:Cif} to also verify that if $\delta_n(x) \equiv 0$, 
then $\delta_{m}(x)=0$ for all $m \geq n$. 

\vskip0.1true in

\noindent Part of AIM's success is due to the inherent simplicity of its 
iterative process governed by the termination condition \eqref{eqI.6} 
as well for its ability to obtain not only explicit  solutions but possibly  
accurate approximations if the solution is not a polynomial. 

\vskip0.1true in

\noindent  Until now,  there is no 
rigorous justification given for the validity of the answers when 
equation  \eqref{eqI.1}  does not have a polynomial solution.   
 In particular the 
treatment of  the constant-coefficient differential equations  in 
\cite{Cif:Hal:Saa} is not only not rigorous but misses an important 
feature of AIM, which we will elaborate on in this work.  There 
were also other cases where AIM failed as pointed out in the works 
of Barakat \cite{Bara}, P. Amore and F. M. Fern\'andez \cite{F2006}, and 
Fernandez \cite{F2004}. Sometimes a transformation of the problem 
makes it amenable for AIM to be applied. 
What is  really mysterious is that AIM appears to give correct answers  
even in certain cases when  $\delta_n(x) \not\equiv 0$ for any finite $n$. There 
are cases 
where AIM did not work, see \cite{F2004} and \cite{Bara} but, besides 
the cases 
of polynomial solutions,  no explanation 
was given as to when AIM works or the iterative procedure in the AIM does not
converge.  
The present paper is a step towards understanding these 
issues. This is a first step towards a rigorous theory   explaining how 
and why AIM works. This is important since AIM is an especially powerful 
tool when it comes to numerical computations.

\vskip0.1true in

\noindent The present work provides not just the mathematical 
justification that accounts for the success of AIM and why the iterative
procedure in the AIM does converge, but we also illustrate cases of failure of the AIM procedure 
in fairly simple examples. Luckily the algorithm which 
implements AIM predicts the cases of failure.  The present work also provides a technique to generate a 
chain of a solvable class of differential equations that approximate the 
exact solution to a given second-order linear differential equation. 
Theorems \ref{thm31}  (see below) is a device which iteratively produces 
new solvable differential equations. 
Theorem \ref{thm31} proves that AIM a robust and reliable technique for 
solving the  differential equations \eqref{eqI.1}. 
\vs We start, in the next section,  by explaining  why AIM gives the correct solutions for a certain  class of  constant-coefficient  differential equations  although, for a fixed $n$, the termination condition \eqref{eqI.6} cannot vanish identically.  Starting from an equation of the form \eqref{eqI.1}, we give a chain of differential equations with two explicitly defined solutions.  The main results of the present work are shown in Sections 2, 3, and 4.  We believe that the case of constant coefficients has all the characteristics of the general case. In the sense that AIM carries the attributes of success and failure as in the general case that depends on the nature of the differential equation coefficients, so we include in  Section 3 a complete treatment of the case of constant coefficients as an illustration of the AIM technique. We show that AIM works if and only if the two roots of the characteristic equation have different moduli unless they are equal.  
We explicitly give examples of linear second-order differential equations with real constant coefficients where the AIM technique fails. Still, luckily, we indicate that the numerics predicts the failure of the method.  
 Section 4 contains all the numerical examples where AIM is implemented and correctly predicts the cases of success or failure. 
 Section 5 treats an example where $\l_0(x)$ is a linear polynomial. This treatment leads to generate a chain of linear second-order differential equations with rational coefficients and explicit bases of solutions.  In Section 6, we give an elementary, but practical, treatment of the classical anharmonic oscillator potential $V(x)=x^2+Ax^4$,  $A\ge 0$. Interestingly, this simple approach gives the eigenvalues accurate to fifty decimals. In Section 7, a conclusion is provided that summarizes the results presented in this work.  
\section{Main Results}\label{MainResults}

\noindent In this section, we state the main results of the present work. 
\begin{thm}\label{thm31}
Let 
\begin{equation}\label{eq2.1}
\begin{gathered}
y_n(x) =  \exp\left(-\int^x \a_n(t) dt\right),\\
 z_n(x)= \exp\left(-\int^x \a_n(t) dt\right)\int^x 
 \exp\left(\int^t [\l_0(u) + 2\a_n(u) du\right) dt.
\end{gathered}
\end{equation}
Then $y_n(x)$ and $z_n(x)$ are linear independent solutions of the differential equation
\begin{align} \label{eq2.2} 
 y^{\prime\prime}(x)-\lambda_0(x)  y^{\prime}(x)&- s_0(x) y(x) = \frac{\l_n(x)s_{n-1}(x) - s_n(x) \l_{n-1}(x)}
 {\l_{n-1}^2(x)}\;   y(x).
\end{align}
\end{thm}

\noindent\textbf{Proof:} It is a calculus exercise to use $y_n^{\prime}(x) = 
-y_n(x) s_{n-1}(x)/\l_{n-1}(x)$ to 
show that $y_n(x)$ satisfies \eqref{eq2.2}.  To find a second 
solution of \eqref{eq2.2} write $y(x) = y_n(x) u_n(x)$ and use the 
variation of parameters to find $u_n(x)$ and the solution $y_n(x)u_n(x)$ 
turns out to be $z_n(x)$.  It can be easily verified that the Wronskian 
of $y_n(x)$ and $z_n(x)$ is not zero.  $\square$
\vs 
Theorem \ref{thm31} explains the gist of AIM. Indeed when 
$\delta_n(x) \equiv 0$ then the perturbed 
differential equation \eqref{eq2.2} 
reduces to \eqref{eqI.1}, so $y_n(x)$ and $z_n(x)$ are two linear 
independent solutions of the original equation \eqref{eqI.1}.  In general  
\eqref{eq2.2} is a perturbation of \eqref{eqI.1}, so what is needed 
is a rigorous theory that shows that for certain class of functions 
$\lambda_0(x)$ and $s_0(x)$ the coefficient of $y(x)$ on the 
right-hand side of \eqref{eq2.2} tends to zero as $n \to \infty$. One 
also needs to show that when $\delta_n(x)/\lambda_0(x)^2 \to 0$ the 
solution of equation \eqref{eq2.2} converge to solution of the 
original equation \eqref{eqI.1}. 
  We shall explore this fact through a number of examples in the next 
  section.
  
\vskip0.1true in

\noindent We note that if $\l_0(x)$ and $s_0(x)$ are rational functions 
then $\l_n(x)$ and $s_n(x)$ are also rational functions, hence 
$\alpha_n(x)$ is a rational function. If we further assume that the poles 
of $\alpha_n(x)$ are all simple then $y_n(x)$, up to  
a constant multiple, takes  the form 
\bea\label{eq2.3} 
y_n(x) = e^{p(x)} \prod_{j}(1-x/x_j)^{a_j},
\eea
where $p(x)$ is a polynomial. On the other hand the function $z_n$ 
involves an indefinite integral of a function of the above form.

\vs
We note that 
\begin{equation} \label{eq2.4}
\begin{gathered}
 \frac{\l_n(x)s_{n-1}(x) - s_n(x) \l_{n-1}(x)} 
 {\l_{n-1}^2(x)} +s_0(x)  =-\dfrac{d}{dx}\left(\dfrac{s_{n-1}(x)}
 {\lambda_{n-1}(x)}\right)+\left(\dfrac{s_{n-1}(x)}
 {\lambda_{n-1}(x)}\right)^2+\lambda_0\left(\dfrac{s_{n-1}(x)}
 {\lambda_{n-1}(x)}\right). 
  \end{gathered}
  \end{equation}
This allows us to write a different form of the differential equation 
\eqref{eq2.2}. 

\section{Constant Coefficients differential equations}\label{Sec3}
\noindent In the case of constant coefficients, the AIM sequences \eqref{eqI.3} reduces to 
  \bea
\label{eq3.1}
\begin{gathered}
\lambda_n  =  
\lambda_{n-1}  \lambda_0  +s_{n-1} ,   \quad
s_n  = \lambda_{n-1} s_0
, \quad n >0.
\end{gathered}
\eea
Therefore,  $\l_n$ and $s_n$ satisfy the recurrence relation 
 \bea
\label{eq3.2}
 \nu_{n+1}= \l_0\, \nu_n + s_0\,\nu_{n-1},
 \eea
 where $\lambda_0$ and $s_0$ are constants. The characteristic 
 equation is   $r^{2}- \l_0 r - s_0=0$ with roots 
\bea
\label{eq3.4}
r_1, r_2 =\dfrac{\l_0}{2}\pm \sqrt{s_0+ \dfrac{\l_0^2}{4}}.
\eea
We may assume $ |r_2| \le |r_1|$. It is clear that  $r_1+r_2=\lambda_0$ and $r_1\,r_2=-s_0.$
\subsection{Two  roots with distinct moduli}
\noindent If    $s_0+\lambda_0^2/4 \ne 0$   then 
\bea \label{eqII.19}
\l_n = Ar_1^n + (\l_0-A)r_2^n, \quad s_n = Br_1^n + (s_0-B)r_2^n,
\eea 
where $A$ and $B$ are  constants,  and $\l_1 = \l_0^2 + s_0, 
s_1 = \l_0s_0$. Also note that $r_1+r_2= \l_0, r_1r_2=-s_0$. 
\bea
\label{eqII.20}
\bg
A(r_1-r_2) = \l_1 -\l_0r_2= \l_0^2+s_0- \l_0r_2 = \l_0r_1+s_0 = r_1^2, \\ B(r_1-r_2)= s_1-s_0=  \l_0s_0 - s_0r_2 = s_0r_1. 
\eg
\eea
Therefore the  ratio $\alpha_{n+1}=s_{n}/\lambda_{n}$ has the form 
\bea\label{eqII.21}
\a_{n+1} = \frac{Br_1^{n} + (s_0-B)r_2^{n}}
{Ar_1^{n} + (\l_0-A)r_2^{n}}.
\eea
In this case $|r_1| >  |r_2|$. Thus, for large $n$, we conclude that  
\bea
\label{eqII.22}
\a_n \approx B/A = s_0/r_1 = -r_2. 
\eea
while
\bea
\label{eqII.23}
\lambda_0+\a_n \approx \lambda_0+B/A = r_1+r_2+s_0/r_1 = r_1, 
\eea
 Equation \eqref{eqI.4} indicates that one solution is $
\exp(-\a_n x)= \exp (r_2\,x)$ while the second linearly independent 
solution is $\exp((\l_0+\a_n)x)=\exp(r_1\, x)$, as expected by the 
classical theory of differential equations with constant coefficients. 
\vskip0.1true in
\noindent It must be noted that in the present case the perturbation term 
in \eqref{eq2.2}  decays exponentially as $n \to \infty$. Indeed
\bea
\notag
\frac{\l_n(x)s_{n-1}(x) - s_n(x) \l_{n-1}(x)}
 {\l_{n-1}^2(x)} = \mathcal{O}(({r_2}/{r_1})^n),
\eea
is true. 
\vskip0.1true in
\noindent In general AIM approach fails when $|r_1|=|r_2|$, and 
$r_1 \ne r_2$. This easily follows from \eqref{eqII.21}. 
The good news is that the AIM algorithm sees this very early on and 
in Section 5 we will illustrate this by an example.

\subsection{Double root}
\noindent In the case of $s_0+\lambda_0^2/4=0$, the solution of the difference equation \eqref{eq3.2} assumes the form, for  $r_1=r_2=r=\lambda_0/2$. Thus $\l_n$ and $s_n$ have the forms
\bea
\label{eqII.24}
 \lambda_{n}= r^n(\l_0 + An), \qquad s_n = r^n(s_0 + Bn). 
 \eea 
with the initial conditions  
 $\lambda_{1}= \l_0^2+s_0, \quad s_1 = s_0\l_0.$ From here 
 it follows that  $A=r, B =-r^2$
\vskip0.1true in
\noindent  The   solution \eqref{eqII.24} now reads
 \bea
\label{eqII.26}
 \lambda_{n}= r^n(n+2), \qquad s_n = -r^{n+2}(n+1).  
 \eea
 Thus, for large $n$,  
\begin{align}\label{eqII.29}
\a_n = s_{n-1}/\lambda_{n-1}  \approx -r. 
\end{align}
Now the general solution reads
\bea
\label{eqII.30}
\bg
y(x) = C_1 e^{rx} + C_2\,x\, e^{rx},
\eg
\eea
again as expected by the general theory of differential equations. 
\vskip0.1true in
\noindent In the present case the perturbation term in Eq.\eqref{eq2.2} is given by
\bea
\notag
\frac{\l_n(x)s_{n-1}(x) - s_n(x) \l_{n-1}(x)}
 {\l_{n-1}^2(x)} = \mathcal{O}(1/n^2),
\eea
confirming that the perturbation term tends to zero. 
 
 We note that matrix form of \eqref{eqI.3}  is 
 \begin{eqnarray}
\label{eqODEls2}
 \begin{gathered}
\left( \begin{matrix}
\lambda_n(x)\\
s_n(x)
\end{matrix} \right) 
= \left[\frac{d}{dx} +  \left( \begin{matrix}
\lambda_0(x) & 1\\
s_0(x) &0
\end{matrix} \right)  \right]
\left( \begin{matrix}
\lambda_{n-1}(x)\\
s_{n-1}(x)
\end{matrix} \right). 
\end{gathered}
\end{eqnarray}
Therefore  the solution is 
\begin{eqnarray}
\label{eqOS}
 \begin{gathered}
\left( \begin{matrix}
\lambda_n(x)\\
s_n(x)
\end{matrix} \right) 
= \left[\frac{d}{dx} +  \left( \begin{matrix}
\lambda_0(x) & 1\\
s_0(x) &0
\end{matrix} \right)  \right]^n
\left( \begin{matrix}
\lambda_{0}(x)\\
s_{0}(x)
\end{matrix} \right). 
\end{gathered}
\end{eqnarray}
This raises the issue of finding an expansion formula for 
\begin{eqnarray}
\label{eqBurchnalltype}
\left[\frac{d}{dx} +  \left( \begin{matrix}
\lambda_0(x) & 1\\
s_0(x) &0
\end{matrix} \right)  \right]^n f(x)
\end{eqnarray}
In the scalar case expansions of the above type are known. For example 
Burchnall \cite{Bur} proved that 
\begin{eqnarray}
\label{eqBur}
\left(\frac{d}{dx}-2x\right)^m f(x) 
=\sum_{k=0}^m(-1)^{n-k}
{m\choose k}H_{m-k}(x)f^{(k)}(x),
\end{eqnarray}
where $H_{m-k}(x)$ is Hermite polynomial. Several related expansions involving Laguerre and Jacobi polynomials, 
as well as their $q$-analogues,  have been established recently in 
\cite{Ism:Koe:Rom} and \cite{Ism:Koe:Rom2}.  We now show how this applies to the case of 
linear second order differential equations with constant coefficients
\vs
 It is easy to see that  
\begin{eqnarray}
 \left( \begin{matrix}
\lambda_0 & 1\\
s_0 &0
\end{matrix} \right) = C D C^{-1}, \quad C = \left( \begin{matrix}
r_1 & r_2 \\
s_{0} & s_{0}
\end{matrix}   \right), D = \left( \begin{matrix}
r_1 & 0\\
0 & r_2
\end{matrix} \right).
\end{eqnarray}
Therefore 
 \begin{eqnarray}
\bg
\left[\frac{d}{dx} +  \left( \begin{matrix}
\lambda_0(x) & 1\\
s_0(x) &0
\end{matrix} \right)  \right]^n f(x) = C  \left(  \begin{matrix}
\exp(-xr_1) & 0\\
0 &\exp(-xr_2)
\end{matrix} \right) \frac{d^n}{dx^n} \left[\left(  \begin{matrix}
\exp(xr_1) & 0\\
0 &\exp(xr_2)
\end{matrix} \right)
  C^{-1}f(x)\right]. 
\eg
\end{eqnarray}

 \noindent Motivated by Burchnall's formula \eqref{eqBur} we formulate the following problem. 
\vs
\noindent{\bf Problem:} Let $\l_0(x)$ and $s_0(x)$ be $n$  times
continuously differentiable functions. Is  there an 
explicit  formula  for 
\bea
\label{eqBurchnalltypeII}
\left[\frac{d}{dx} +  \left( \begin{matrix}
\lambda_0(x) & 1\\
s_0(x) &0
\end{matrix} \right)  \right]^n \left( \begin{matrix}
f_1(x)\\
f_2(x)
\end{matrix} \right). 
\eea

  
\section{Applications: Constant coefficient differential equations}

\begin{figure}[!h]
\centering
\includegraphics[height=2in, width=2.5in]{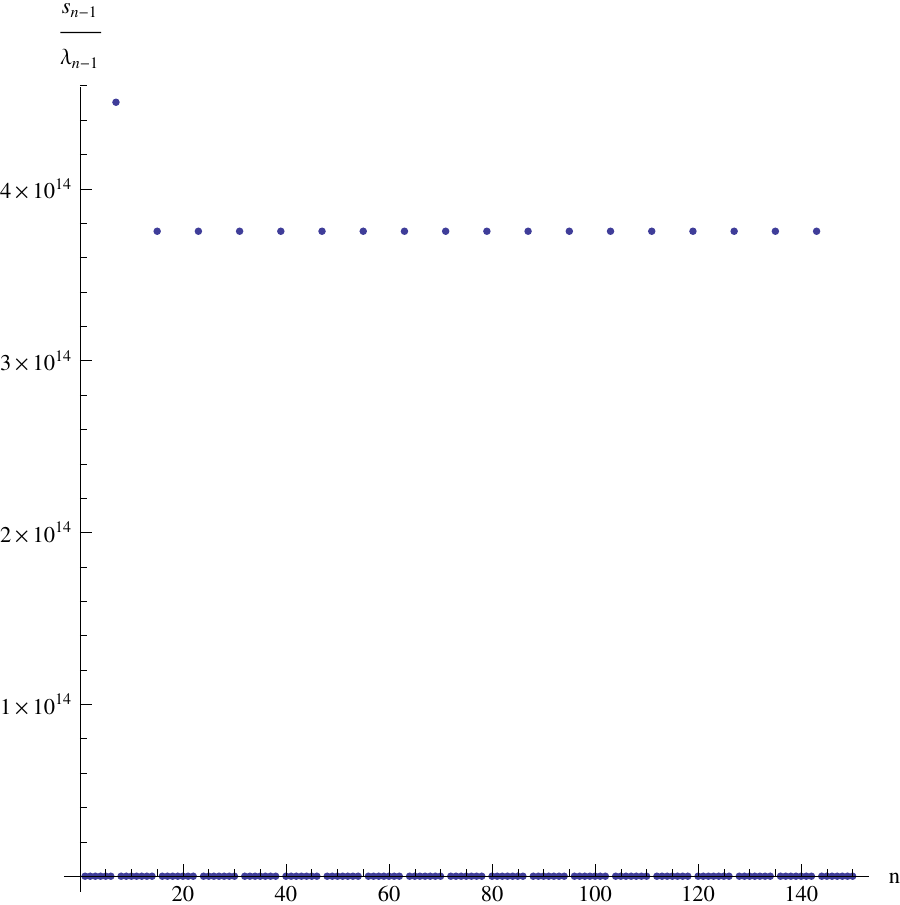}\hskip0.5true in\includegraphics[height=2in, width=2.5in]{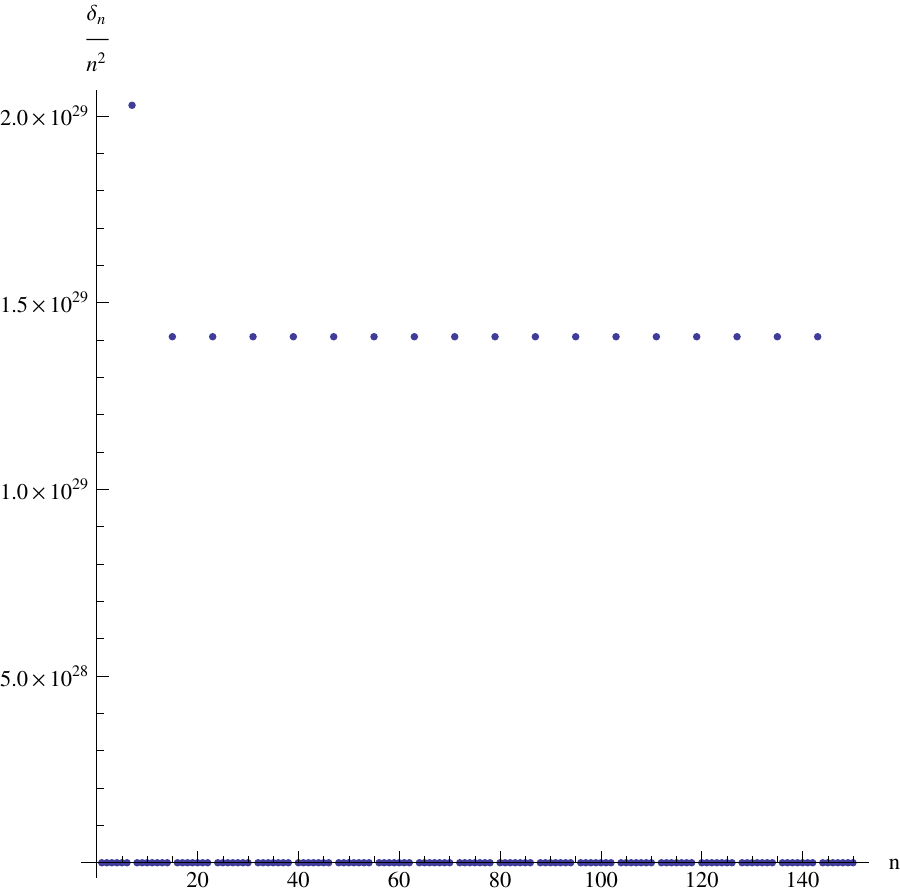}
\caption{The discrete points along the horizontal line and the $n$-axis illustrate the oscillation behaviour of $\alpha_n$ and $\delta_n/\lambda_{n-1}$ for $n=1,2,\dots,150$ produced by AIM in handling the differential equation
 \eqref{eq5.1}. }
\label{Complex}
\end{figure}

\noindent  In the original publication of AIM \cite{Cif:Hal:Saa}, the authors explore its application in solving the constant coefficient differential equations when the roots of the associated characteristic equation are real distinct. No attempt for the other possible cases was ever discussed. 
This may have gave a misleading impression that AIM always applicable for this class of equations. The examination for AIM convergence, Section \ref{Sec3},  shows a drawback in AIM usage in handling the constant coefficient differential equations when the associated characteristic equations has `equal\rq{} moduli complex roots. Although, this is not a small class of equations, the information extracted from AIM usage are not all lost, for the oscillation behavior of $\alpha_n$ serve as an indicator for the peculiar form of the complex roots. 
Consider, for example, the differential equation
\begin{align}\label{eq5.1}
y''-2\cos(\pi/8) y'+\,y=0,
\end{align}
The evaluation of $\alpha_n=s_{n-1}/\lambda_{n-1},$ and $\delta_n(x)/\lambda_{n-1}^2$ for $ n=1,2,\dots,150$  are illustrated in the figure \ref{Complex}.  

\begin{figure}[h]
\centering
\includegraphics[height=2in, width=2.5in]{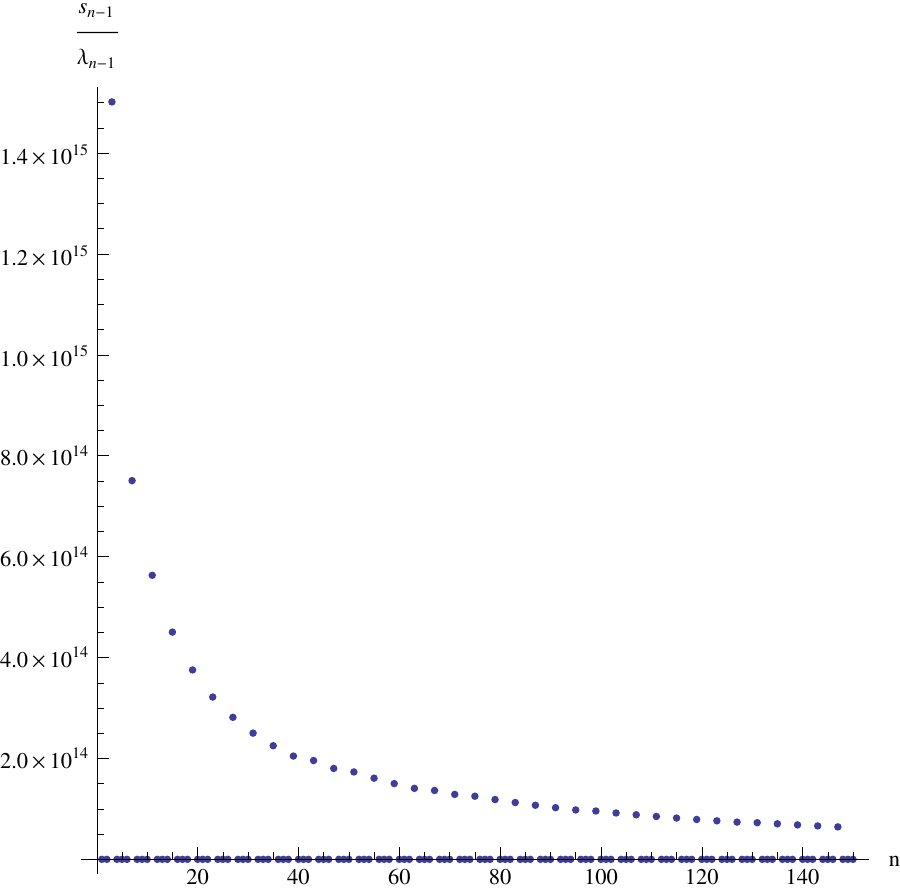}\hskip0.75true in\includegraphics[height=2in, width=2.5in]{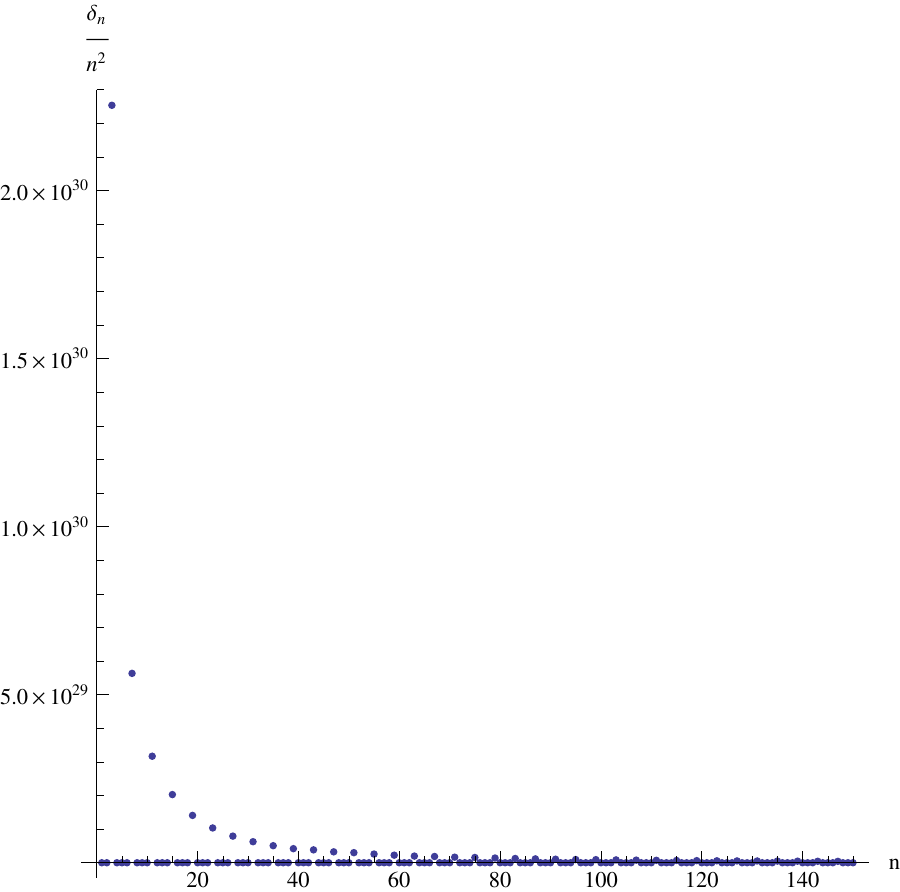}
\caption{The discrete points along the exponential-type curve and the $n$-axis illustrate the strange behaviour of $\alpha_n$ and $\delta_n/\lambda_{n-1}$ for $n=1,2,\dots,150$ produced by AIM in handling the differential equation \eqref{eq5.2}.}
\label{Complex2}
\end{figure}

\noindent This oscillation behavior predict that the roots of the characteristic equation
$
r^2-2\cos(\pi/8) r+\,1=0
$, namely, $r_1=\cos(\pi/8)+i\sin(\pi/8)$ and $r_2=\cos(\pi/8)-i\sin(\pi/8)$ have equal moduli.  We illustrate another strange behaviour produced by AIM numerical evaluation by the following example.
Consider  the differential equation
\begin{align}\label{eq5.2}
y''-2\cos(\pi/4) y'+\,y=0,
\end{align}
The evaluation of $\alpha_n$ and $\delta_n(x)/\lambda_{n-1}$ using AIM sequences are displayed in Figure  \eqref{Complex2}.
\noindent Again such response of AIM is confirmed by the modulus of 
the complex roots of the characteristic equation, namely 
$r_1=(1 - i)/\sqrt{2}$ and $r_2=(1+i)/\sqrt{2}$.  

Let us now consider the 
general case when $\l_0$ and $s_0$ are real and $|r_1|=|r_2| = r$, 
say. Let $r_1, r_2 = r e^{i\t}, 
re^{-i\t}$. In this case \eqref{eqII.21} gives 
\bea
\notag
\alpha_{n+1} = \frac{2\,i\,B\, \sin (n\,\t) + s_0\, e^{-in\t}}
{2\,i\,A\, \sin (n\,\t) + \l_0\, e^{-in\t}}.
\eea
If $\t$ is a rational number $\in (0, \pi/2)$ we can choose $n$ to be an 
arithmetic  progression which makes $\sin (n\t) =0$ while $r_1-r_2 
= 2 i r \sin \t \ne 0$, hence $A$ and $B$ are well defined. On such a 
sequence $\alpha_{n+1}$ takes the constant value $s_0/\l_0$, which gives the wrong answer. The point we are trying to make is that in such cases AIM fails and looking at subsequences does not help. 

\begin{figure}[h]
\centering
\includegraphics[height=2in, width=2.5in]{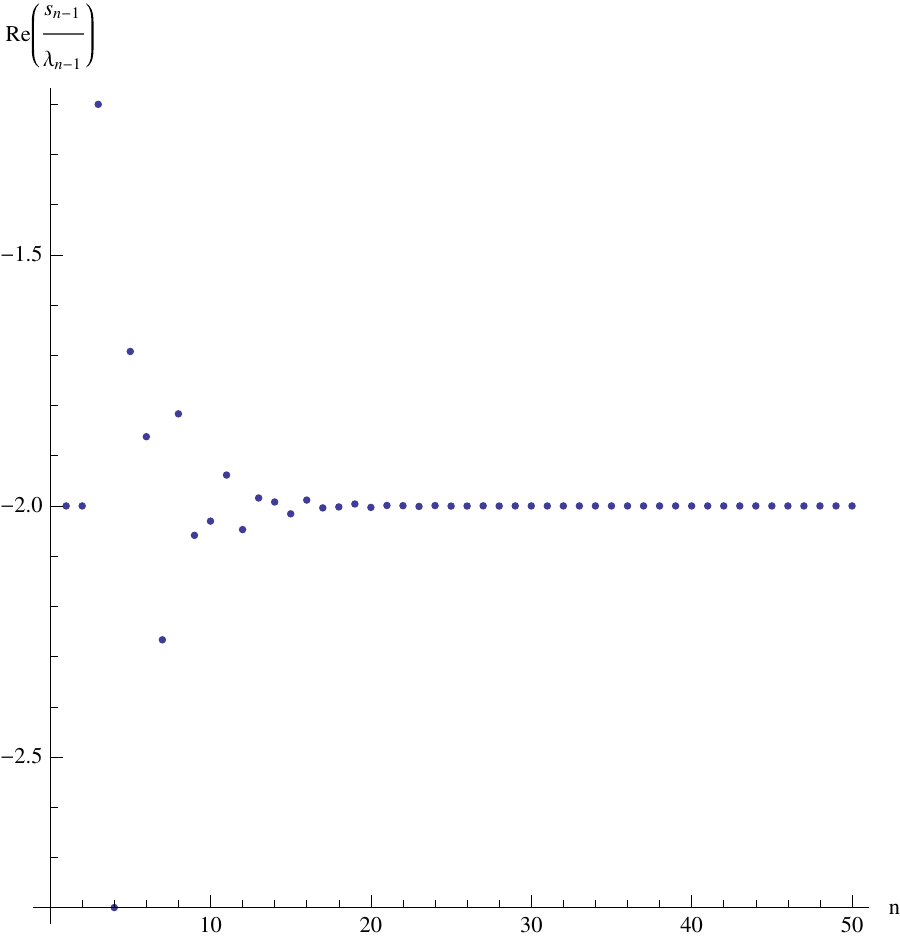}
\hskip0.1true in\includegraphics[height=2in, width=2.5in]{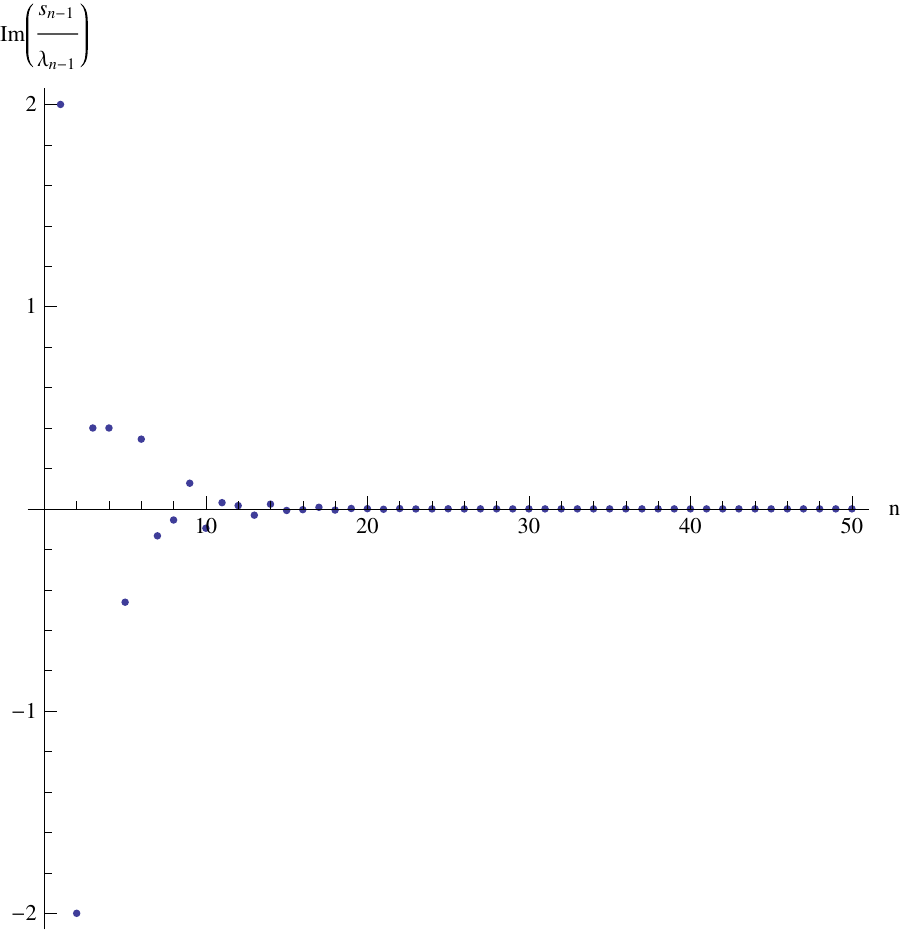}\\\includegraphics[height=2in, width=2.5in]{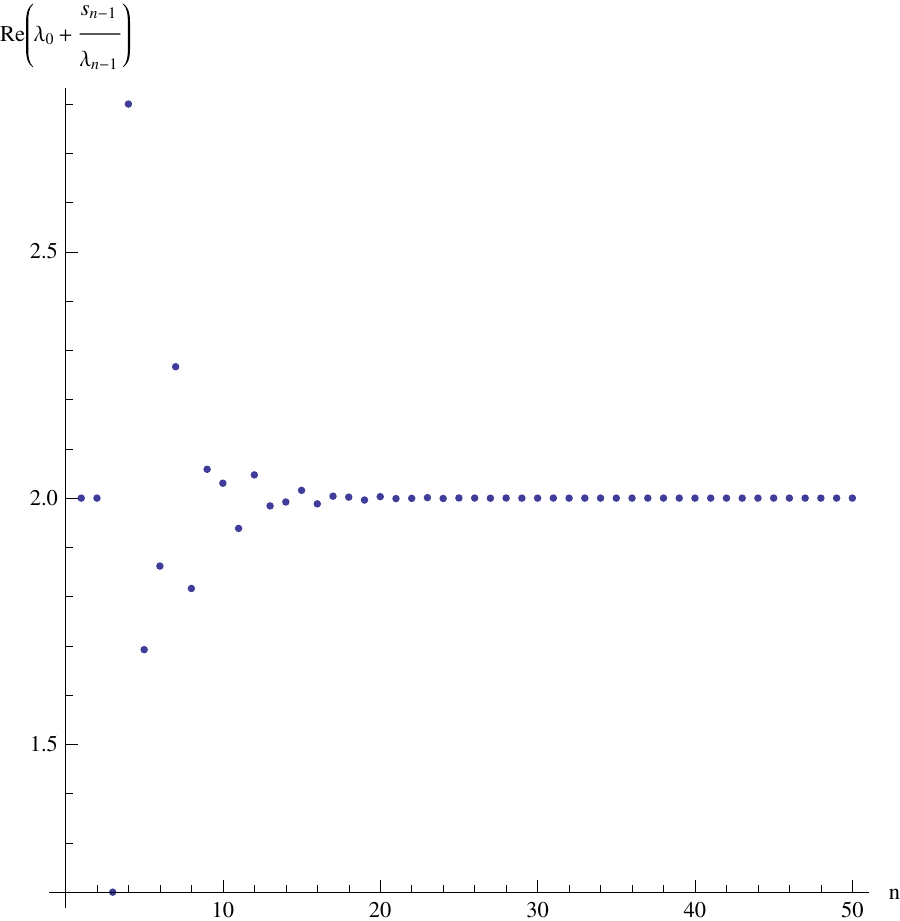}\hskip0.1true in\includegraphics[height=2in, width=2.5in]{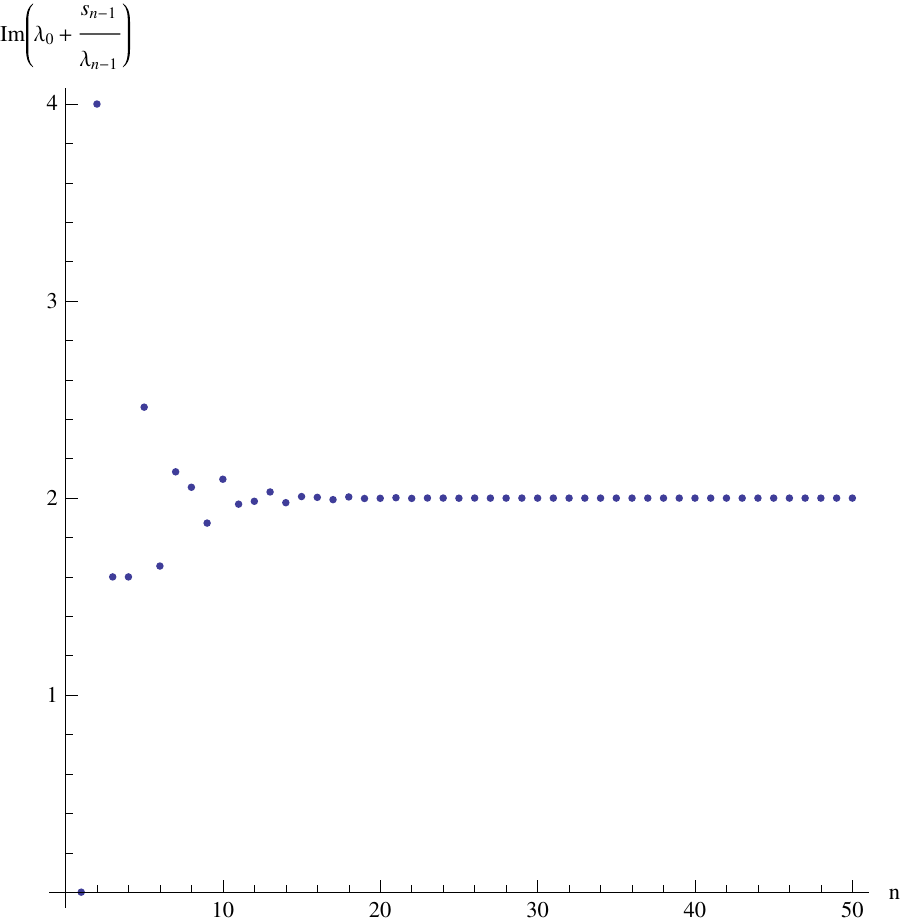}\caption{The convergence of $\alpha_n$ and $\lambda_0+\alpha_n$ for $n=1,2,\dots,50$, produced by AIM, to exact solutions of the equation \eqref{eq5.4}.}
\label{Complex3}
\end{figure}

Another class that was overlooked in \cite{Cif:Hal:Saa} is the  linear differential equations with complex coefficients. AIM gives the exact solutions as long as the roots of the associated characteristic equation have different norms. A class of such equations is given by
\begin{align}\label{eq5.3}
y''-a\,i\, y'-b\,(b+a\,i)\,y=0,\qquad i=\sqrt{-1},\quad a,b\in \mathbb R
\end{align}
Take $a=b=2$, Eq.\eqref{eq5.3} then reads
\begin{equation}\label{eq5.4}
y''(x)-2\,i\,y'(x)-(4+4\,i)\,y(x)=0.
\end{equation}
AIM tends to the exact solutions $\{ e^{2x}, e^{(2+2i)x}\}$ as $n$ gets larger and larger as the figure \eqref{Complex3} shows.

\section{Applications: Rational coefficient differential equations}
\noindent As an  example, we establish classes of exact-solvable differential equations initiated with rational coefficients in the form
\begin{equation}\label{eqV.47}
y''(x)-\mu\,x\,y'(x)+\eta\,y(x)=0.
\end{equation}
Starting with $\lambda_0=\mu\,x$ and $s_0=-\eta,$  the first-iteration using Theorem \ref{thm31} implies
\begin{equation}\label{eqV.48}
y_1''(x)-\mu\,x\,y_1'(x)+\eta\, y_1(x)=\frac{\eta (\eta-\mu)}{\mu^2 x^2}\,y_1(x)~\mbox{with}~ 
y_1(x)=x^{\eta/\mu}.
\end{equation}
This differential equation generate, for example,  infinite classes of exactly solvable differential equations
\begin{equation}\label{eqV.49}
x^2u''(x)-\mu\,x^3\,u'(x)+m\left(\mu\, x^2 -m+1\right)\,u(x)=0\quad\mbox{with}\quad 
u(x)=x^{m},~m=0,\pm 1,\pm 2,\dots,
\end{equation}
and, with the replacement $x$ with $1/x$,
\begin{equation}\label{eqV.50}
x^4\,v''(x)+x\left(2\,x^2+\mu\right)\,v'(x)+m\left(\mu -(m-1)\,{x^2}\right)\,v(x)=0~\mbox{with}~~ 
v(x)={x^{-m}},
\end{equation}
for $m=0,\pm 1,\pm 2,\dots$. The second iteration, using  Theorem \ref{thm31}, yields
\begin{equation}\label{eqV.51}
y_2''(x)-\mu\,x\,y_2'(x)+\eta\, y_2(x)=\frac{\eta(\eta-\mu) (\eta-2\mu)}{\left(\mu-\eta+\mu^2 x^2\right)^2}\,y_2(x)\quad \mbox{with}\quad 
y_2(x)=\left(\eta-\mu -\mu^2 x^2\right)^{{\eta}/({2\mu})}.
\end{equation}
This differential equation also generate  infinite classes of exactly solvable differential equations, for example,
\begin{align*}
y_2''(x)-\mu\, x\, y_2'(x)+2\, m\, \mu\, \left(1-\frac{2 (m-1) (2 m-1)}{\left(1-2 m+\mu x^2\right)^2}\right) y_2(x)=0, 
\end{align*}
where
\begin{align*}
y_2(x)=(1-2m+\mu \,x^2)^m,\quad m=0,1,2,\dots.\end{align*}
The third iteration, using  Theorem \ref{thm31}, yields
\begin{equation}\label{eqV.52}
y_3''(x)-\mu\,x\,y_3'(x)+\eta \, y_3(x)=\frac{\eta (\eta-\mu) (\eta-2\mu) (\eta-3\mu)}{\mu^2 x^2 
\left(3\mu-2 \eta+\mu^2 x^2\right)^2}\,y_3(x)
\end{equation}
with exact solution 
\begin{equation}\label{eqV.53}
y_3(x)=x^{{\eta (\eta-2 \mu)}/({(2 \eta-3\mu) \mu})}\left(2\eta-3\mu -\mu^2 x^2\right)^{{\eta(\eta-\mu)}/{(2\mu({2 \eta-3 \mu})})}\end{equation}
The fourth iteration, using  Theorem \ref{thm31}, yields
\begin{equation}\label{eqV.54}
y_4''(x)-\mu\,x\,y_4'(x)+\eta\, y_4(x)=\frac{\eta (\eta-\mu) (\eta-2\mu) (\eta-3\mu)(\eta-4\mu)}
{(\eta-3\mu) (\eta-\mu)-3 (\eta-2 \mu) \mu^2 x^2+\mu^4 x^4}\,y_4(x)
\end{equation}
with exact solution given by
\begin{align}\label{eqV.55}
y_4(x)&=\left((\eta-3 \mu) (\eta-\mu)-3 (\eta-2 \mu) \mu^2 x^2+\mu^4 x^4\right)^{{\eta}/({4\mu})}\notag\\
&\times\left(\sqrt{5\eta^2-20 \eta \mu+24 \mu^2}+ 3 \eta-2 \mu \left(3+\mu x^2\right)\right)^{-\tfrac{\eta (\eta-4 \mu)}{4 \mu \sqrt{5 \eta^2-20 \eta \mu+24 \mu^2}}}\notag\\
&\times \left({\sqrt{5 \eta^2-20 \eta \mu+24 \mu^2}}-3 \eta+2 \mu \left(3+\mu x^2\right)\right)^{\tfrac{\eta (\eta-4 \mu)}{4 \mu \sqrt{5 \eta^2-20 \eta \mu+24 \mu^2}}}\end{align}
The fifth iteration, using  Theorem \ref{thm31}, yields
\begin{align}\label{eqV.56}
y_5''(x)-\mu\,x\,y_5'(x)&+\eta\, y_5(x)=\dfrac{\eta (\eta-\mu) (\eta-2\mu) (\eta-3\mu)(\eta-4\mu)(\eta-5\mu)}
{\mu^2 x^2 \left(3 \left(\eta^2-5 \eta \mu+5 \mu^2\right)-2 (2 \eta-5 \mu) \mu^2 x^2+\mu^4 x^4\right)^2}\,y_5(x)
\end{align}
with exact solution given by
\begin{align}\label{eqV.57}
y_5(x)&=x^{\tfrac{\eta (\eta-4 \mu) (\eta-2 \mu)}{3\mu \left(\eta^2-5 \eta \mu+5 \mu^2\right)}}\notag\\
&\times\left(5\mu-2 \eta+\mu^2 x^2-\sqrt{\eta^2-5\eta \mu+10 \mu^2}\right)^{\tfrac{\eta (\eta-\mu) \left((2\eta-7 \mu) \sqrt{\eta^2-5\eta \mu+10 \mu^2}-(\eta-5 \mu) (\eta-4 \mu)\right)}{12\mu \left(\eta^2-5 \eta \mu+5 \mu^2\right) \sqrt{\eta^2-5 \eta \mu+10 \mu^2}}}\notag\\
&\times\left(5\mu -2 \eta+\mu^2 x^2+\sqrt{\eta^2-5 \eta \mu+10 \mu^2}\right)^{\tfrac{\eta (\eta-\mu) \left((\eta-5\mu) (\eta-4\mu)+(2\eta-7\mu) \sqrt{\eta^2-5 \eta \mu+10 \mu^2}\right)}{12 \mu \left(\eta^2-5 \eta \mu+5 \mu^2\right) \sqrt{\eta^2-5\eta \mu+10 \mu^2}}}
\end{align}
Clearly, for $\eta=m \mu$, $m=0,1,2,\dots$, the general solutions of the generalized Hermite equation
\begin{equation}\label{eqV.58}
y_m''(x)-\mu\,x\,y_m'(x)+m\,\mu\, y_m(x)=0,\quad m=0,1,2,\dots.
\end{equation}
reads
\begin{align}\label{eqV.59}
y_0(x)&=1,\quad y_1(x)=x,\quad y_2(x)=\mu\,x^2-1,\quad y_3(x)=x(\mu x^2-3),~\dots
\end{align}
with a general expression given in terms of the Hermite polynomials as
\begin{equation}\label{eqV.60}
y_m:= H_{m}\left(\sqrt{\dfrac{\mu}{2}}\,x\right),
\end{equation}
up to a constant. 

\section{One-dimensional anharmonic oscillator potentials}
\noindent We consider a class of one-dimensional anharmonic oscillator potentials discussed by Ciftci, Hall and Saad in their original work on AIM that reflects on its  powerful computaional aspect:
\begin{align}\label{eq7.1}
V(x)= x^2+A\,x^4,\quad A\geq 0,\quad x\in(-\infty,\infty).
\end{align}
Schr\"odinger's equation for $V(x)$ takes the form  (in the units $m=2\hbar =1$)
\begin{align}\label{eq7.2}
-\dfrac{d^2\psi}{dx^2}+\left(x^2+A \,x^4\right)\psi=E\psi,\quad \int_{-\infty}^\infty |\psi(x)|dx<\infty.
\end{align}
Using the asymptotic solution near $A=0,$ a general solution of \eqref{eq7.2} take the form
\begin{align}\label{eq7.3}
\psi(x)=e^{-x^2/2}f(x),
\end{align} 
which up on substituting into \eqref{eq7.2} gives the differential  equation of $f(x)$ as
\begin{align}\label{eq7.4}
\dfrac{d^2f}{dx^2}-2\,x\, f'(x)-\left(1-E+A\, x^4
\right)\,f=0.
\end{align}
Theorem \ref{thm31} gives, as a first approximation of equation \eqref{eq7.4},
\begin{align}\label{eq7.5}
f''(x)-2\,x\, f'(x)&-\left(1-E+A\, x^4
\right)\,f(x)=\frac{(E-3) (E-1)-2 A (E+2)\, x^4+A^2\, x^8}{4\, x^2}f(x),
\end{align}
with exact solution given by
\begin{align}\label{eq7.6}
f(x)=x^{(E-1)/2}e^{-A\,x^4/8}.
\end{align}
The right-hand side of \eqref{eq7.5} confirm that the termination condition $\delta_1(x)=0$ depends explicitly on the variable $x\neq 0$ and the unknown $E$ for the given parameter $A$.  Further, for the initial starting of $x_0$, we should start sighlty away from $x_0=0$ because $\delta_1(x)=0$ will yields $E=1$ or $E=3$. Thus,  for a good initial starting of $x_0$ is $x_0=0.0001$.
As a second approximation, Theorem \ref{thm31} gives
\begin{align*}
f''(x)-2\,x\, f'(x)&-\left(1-E+A\, x^4
\right)\,f(x)=\tfrac{(E-5) (E-3) (E-1)+12 A (E-3)\, x^2-3 A \left(5+2E+E^2\right) x^4+20 A^2 x^6+3 A^2 (E+5) x^8-A^3 x^{12}}{\left(3-E+4 x^2+A x^4\right)^2}f(x)
\end{align*}
with exact solution
\begin{align*}
f(x)&=e^{-x^2}\left(3-E+4 x^2+A x^4\right)^{(2/A)-1}\left(1-\tfrac{\left(2+A x^2\right)}{\sqrt{4+A (E-3)}}\right)^{\tfrac{3 A-4}{A \sqrt{4+A (E-3)}}}
\left(1+\tfrac{\left(2+A x^2\right)}{\sqrt{4+A (E-3)}}\right)^{-\tfrac{(3 A-4)}{A \sqrt{4+A (E-3)}}}.
\end{align*}
As a third approximation, Theorem \ref{thm31} gives
\begin{align*}
\dfrac{d^2f}{dx^2}-2\,x\, f'(x)-\left(1-E+A\, x^4
\right)\,f&=\tfrac{(x-7)(x-5)(x-3)(x-1)+ 48 A \left(1-5E+E^2\right) x^2+ 4 A \left(60 A-11E-E^3\right) x^4}{16 x^2 \left(E-\left(2+x^2\right) \left(2+A x^2\right)\right)^2}\\
&+\tfrac{16 A^2 (31+2E) x^6+6 A^2 \left(25+8E+E^2\right) x^8-80 A^3 x^{10}-4 A^3 (8+E) x^{12}+A^4 x^{16}}{16 x^2 \left(E-\left(2+x^2\right) \left(2+A x^2\right)\right)^2}
\end{align*}
with exact solution
\begin{align*}
f(x)&=x^{{(E-5) (E-1)}/({4 (E-4)})}\,e^{-\left(4-4A+Ax^2\right)x^2/{16}} \notag\\
&\times\left(1+A+Ax^2-\sqrt{(1-A)^2+AE}
\right)^{\frac{16-45 A+19 A^2-16 A^3+4 \left(A^3-A^2+3 A-1\right)E+\left(16A^2+35 A-16-4 \left(A^2+2A-1\right) E\right) \sqrt{1+A (A-2+E)}}{16 A (E-4) \sqrt{1+A (A-2+E)}}}\notag\\
&\times \left(1+A+Ax^2+\sqrt{(1-A)^2+AE}
\right)^{-\frac{16-45 A+19 A^2-16 A^3-4(1-3 A+A^2- A^3)E+\left(16-35 A-16 A^2+(4A^2+8A-4)E\right) \sqrt{1+A (A-2+E)}}{16 A (E-4) \sqrt{1+A (A-2+E)}}}
\end{align*}

\noindent Although for higher iteration numbers the explicit expressions for $\delta_n(x)/\lambda_{n-1}^2$, using theorem \ref{thm31}, become tedious to handle, they are quite manageable using any Computer Algebra Systems, such as `Maple' or `Mathematica.' 
\vskip0.1true in
\noindent In Table \ref{Table1}, the computed eigenvalues for the first fifty significant figures are reported. These stated eigenenergies are more accurate than that reported, for instant, by Pedram et al. \cite{Ped}. For higher values of the parameter $A$, AIM still valuable to handle it, in the case of $A=2$, for example, AIM gives $E_0=1.607~541~302~469$ with $368$ iteration without the requirements of highly sophisticated mathematical analysis \cite{B1973,T2005}.

\begin{table}[!h]
\begin{center}
\begin{tabular}{ |c|c|} 
 \hline 
$n$ & $E_n$ \\ \hline &  \\[-1em]
0 & $1.065~285~509~543~717~688~856~877~962~022~551~287~191~163~282~841~44_{152, 45.614}$  \\ &  \\[-1em]  \hline &   \\[-1em]
$1$ & $3.306~872~013~152~913~507~126~866~993~202~085~609~482~310~246~676~21_{159, 342.010}$  \\ &  \\[-1em]  \hline &   \\[-1em] 
$2$ & $5.747~959~268~833~563~304~734~474~846~968~694~805~582~344~997~674~23_{ 176, 439.805} $  \\ &  \\[-1em]  \hline &   \\[-1em] 
$3$ & $8.352~677~825~785~754~712~154~419~082~681~400~254~841~719~288~378~95_{ 193, 578.345}$ \\ &  \\[-1em]  \hline &   \\[-1em] 
$4$ & $11.098~595~622~633~043~011~086~729~362~490~999~051~260~587~293~654~68_{ 216, 1004.121}$\\ &  \\[-1em]  \hline &   \\[-1em] $5$ & $13.969~926~197~742~799~300~973~142~589~965~547~699~648~595~880~717~04_{ 233, 1151.930}$\\ &  \\[-1em]  \hline &   \\[-1em] $6$ & $16.954~794~686~144~151~337~691~635~822~446~656~334~250~292~275~100~95_{240, 1003.714}$\\ &  \\[-1em]  \hline &   \\[-1em] 
$7$ & $20.043~863~604~188~461~233~636~831~676~507~839~006~262~656~766~921~28_{247, 1635.884}$\\ &  \\[-1em]  \hline &   \\[-1em] $8$ & $23.229~552~179~939~289~070~647~028~684~346~580~007~513~693~999~499~08_{280, 1676.958}$\\ &  \\[-1em]  \hline $9$ & $26.505~554~752~536~617~417~468~366~858~512~851~536~033~849~697~560~07_{287, 1967.322}$\\   \hline
\end{tabular}\vs
\caption{The first ten energy levels of the anharmonic oscillator described by the Hamiltonian $-d^2/dx^2+x^2+0.1\, x^4$. Subscript numbers refer to the number of iterations followed by the time (in seconds).}
\label{Table1}
\end{center}
\end{table}

\section{Conclusion}\label{conc}

\noindent The mystery of AIM in handling certain eigenvalue problems while the breakdown in working with others has been a source of frustration for many researchers  
  \cite{F2004,F2006,K2007}. In most of the published work on AIM, this failure was marked by the oscillation problem associated with the numerical computation of the termination condition $\delta_n(x)$ as $n$ increased. Certain cautions have been devised to handle such oscillation behavior in some cases \cite{Bara,B2008,saad2011,Ri:Sa:Se,Ri:NS,cho:cor:dou,cho:dou:nay}.  Theorem \eqref{thm31} serve as an indicator to measure AIM's success: \emph{for AIM to work, it is necessary that
\begin{equation}\label{eq56}
\left|\dfrac{\delta_{n+2}(x)}{\lambda_{n+1}^2}-\dfrac{\delta_{n+1}(x)}{\lambda_{n}^2}\right|\to 0, \qquad \mbox{for large}\quad  n.
\end{equation}}

\section{Acknowledgments and Funding}
\medskip
\noindent Partial financial support of this work under Grant No. GP249507 from the
Natural Sciences and Engineering Research Council of Canada
 is gratefully acknowledged. A working Mathematica program is available up on request.


\end{document}